# EXACT AND APPROXIMATE ANALYTIC SOLUTIONS
# IN THE SIR EPIDEMIC MODEL

Mário Berberan-Santos
Institute for Bioengineering and Biosciences
Instituto Superior Técnico
Universidade de Lisboa
1049-001 Lisboa
Portugal

berberan@tecnico.ulisboa.pt

**Abstract**

In this work, some new exact and approximate analytical solutions are obtained for the SIR epidemic model, which is formulated in terms of dimensionless variables and parameters, reducing the number of independent parameters from 4 ($I_0$, $S_0$, $\beta$, $\alpha$) to 2 ($i_0 = I_0/S_0$ and $R_0 = \beta S_0/\alpha$). The susceptibles population is in this way explicitly related to the infectives population using the Lambert W function (both the principal and the secondary branches). A simple and accurate relation for the fraction of the population that does not catch the disease is also obtained. The explicit time dependences of the susceptibles, infectives and removed populations, as well as that of the epidemic curve are also modelled with good accuracy for any value of $R_0$ using simple functions that are modified solutions of the $R_0 \to \infty$ limiting case (logistic curve). It is also shown that for small $i_0$ ($i_0 < 10^{-2}$) the effect of a change in this parameter on the population evolution curves amounts to a time shift, their shape and relative position being unaffected.

**Highlights**

- The susceptibles population is explicitly related to the infectives population using the Lambert W function (both principal and secondary branches).

- The time dependence of the susceptibles, infectives and removed populations, as well as that of the epidemic curve are described with good accuracy using simple functions.

- A change in the initial number of infectives on the population evolution curves is shown to amount to a time shift (provided the initial number of infectives is much smaller than that of the susceptibles).

## 1. Introduction

The propagation of new viral infectious diseases can be described, to a first approximation, according to the SIR model developed by Kermack and McKendrick [1-6]. In this deterministic compartmental model, three classes of individuals are considered: Susceptibles, *S*, who can catch the disease; Infectives (infected and infectious), *I*, who can transmit the disease to susceptibles; and Removed, *R*, who are former infectives that either recovered from the disease and became immune to it (permanently or in the time frame of the analysis) or died because of the disease. It is usually assumed that no members of the Removed class are initially present. Given the timescale of a typical epidemic, birth, death (by other causes) and other demographic processes can usually be neglected. It is further assumed that the latency period is negligible, implying that an individual, once infected, immediately becomes infective. The model also assumes a perfect mixing of Susceptibles and Infectives, as a result of free motion (extensive interconnectedness) within the closed system. With these assumptions, the model can be written as the following pair of evolution steps:

$$S + I \xrightarrow{\beta} 2I \qquad (1)$$

$$I \xrightarrow{\alpha} R \qquad (2)$$

The first step accounts for the infection process that occurs via close contact of an infective with a susceptible, and $\beta$ is the infection rate constant. It reflects the intrinsic frequency of contacts and the probability of transmission upon contact: according to the theory of diffusion-influenced collisional processes, it can be written as $\beta = pc$, where *c* is the encounter rate for full diffusion control and *p* is the probability of transmission upon encounter [6]. The second step describes the immunization or death of infectives that thus become members of the Removed class, and $\alpha$ is the removal rate constant, representing the probability of transition $I \to R$ per unit time.

The nonlinear system of differential equations corresponding to the SIR mechanism is [1]:

$$\frac{dS}{dt} = -\beta S I, \tag{3}$$

$$\frac{dI}{dt} = \beta S I - \alpha I = \alpha I \left( \frac{\beta S}{\alpha} - 1 \right), \tag{4}$$

$$\frac{dR}{dt} = \alpha I, \tag{5}$$

where $S$, $I$ and $R$ are population densities (number of class elements per unit area). As $R$ does not appear in the first two equations, the problem is effectively reduced to a system of two (autonomous) differential equations. The dynamics is determined by four parameters: (i) Initial number of susceptibles $S_0$; (ii) Initial number of infectives $I_0$ (assumed to be suddenly mixed with the susceptibles at time zero). $I_0$ is usually much smaller than $S_0$ (it can be as low as a single individual in the entire system); (iii) Rate constant $\beta$ that reflects the likeliness of disease transmission upon contact between an infective and a susceptible; (iv) Rate constant $\alpha$ that reflects the average time needed for an infective to be removed. The inverse of $\alpha$ is the average infectious period $\tau$ [2-5], which is also the average duration of an infective. Another quantity of interest is the epidemic curve, $C$ [2-5]

$$C = \beta S I, \tag{6}$$

which is the number of new infectives per unit time, as follows from equation (4).
The purpose of this work is to obtain new explicit relations for the SIR model (formulated in terms of dimensionless variables) both exact and approximate, including simple analytic solutions for the time-dependence that are valid for any $R_0$. The paper is organized as follows: In Section 2, the SIR model is formulated in terms of dimensionless variables and the susceptibles population is explicitly related to the infectives population using both the Lambert W function and an approximate yet accurate function (sum of exponentials). The maximum value of the epidemic curve is also explicitly obtained. In Section 3, the explicit time dependence of the susceptibles, eq. (38), and infectives, eq. (41), as well as that of the epidemic curve, eq. (45), are obtained in terms of approximate analytic functions whose respective parameters are obtained in Section 4 as a function of $R_0$. The introduced functions also allow to define analytically the respective maximum times. In Section 5, the effect of $i_0$ ($i_0 \ll 1$) on the time-dependence is shown to amount to a time shift. Finally, in Section 6 the main results are summarized.

## 2. Results

Upon application of Laplace transforms, it follows from equations (3) and (4) that

$$I = I_0 \exp\left(-\frac{t}{\tau}\right) + C \otimes \exp\left(-\frac{t}{\tau}\right), \tag{7}$$

$$R = \alpha I \otimes 1 = I_0 - I + C \otimes 1 = S_0 + I_0 - (S + I), \tag{8}$$

where $\otimes$ stands for the convolution between two functions. The epidemic curve thus acts as the impulse function that defines the infectives time evolution.

The system of differential equations can be rewritten in terms of five reduced (dimensionless) quantities:

$$s = \frac{S}{S_0}, \tag{9}$$

$$i = \frac{I}{S_0}, \tag{10}$$

$$r = \frac{R}{S_0}, \tag{11}$$

$$R_0 = \frac{\beta S_0}{\alpha} = \beta S_0 \tau, \tag{12}$$

$$\theta = \alpha t, \tag{13}$$

where $R_0$ is the basic reproduction number of the infection [2-6]. This quintet is preferable to another set, not entirely dimensionless, previously proposed [7].

Equations (3)-(5) become:

$$\frac{ds}{d\theta} = -R_0 s i, \tag{14}$$

$$\frac{di}{d\theta} = (R_0 s - 1) i, \quad (15)$$

$$\frac{dr}{d\theta} = i. \quad (16)$$

The initial values of $s$ and $i$ are 1 and $i_0$, respectively. Usually, $i_0 \ll 1$.

The reduced epidemic curve, $c$, is

$$c = R_0 s i, \quad (17)$$

In this way, the evolution of the system is fully defined by a single initial condition, $i_0$, and by a single dynamic parameter, $R_0$. This parameter has the meaning of the average number of susceptibles infected by a single infective at the beginning of the epidemic [2-7]. Values up to 15 are known [8], while for COVID-19 in pre-pandemic conditions it was between 2.4 and 2.5 [9,10]. The $R_0$ value reflects not only the intrinsic aspects of the disease, but also the population specific conditions (frequency of contacts, probability of transmission).

The threshold (or critical) number of susceptibles, is

$$s_c = \frac{1}{R_0}. \quad (18)$$

If $R_0 < 1$ ($s_c > 1$), the number of infectives monotonically decreases with time (cf. equation (15)) and the epidemic cannot proceed ('herd immunity' situation) [1-6]. On the other hand, if $R_0 > 1$ ($s_c < 1$), the epidemic ensues, with the number of infectives increasing until $s_c$ is reached.

Owing to the nonlinear terms, no explicit analytical solution exists for the system of differential equations (14)-(16), and the time evolution must be obtained numerically or using approximate equations. A few analytical results are nevertheless possible.

Elimination of time from equations (14) and (15) yields

$$\frac{di}{ds} = \frac{1}{R_0 s} - 1, \quad (19)$$

whose integration gives the (phase space) relation between *i* and *s* [1-6]:

$$i = 1 + i_0 - s + \frac{1}{R_0} \ln s, \tag{20}$$

depicted in Fig. 1.

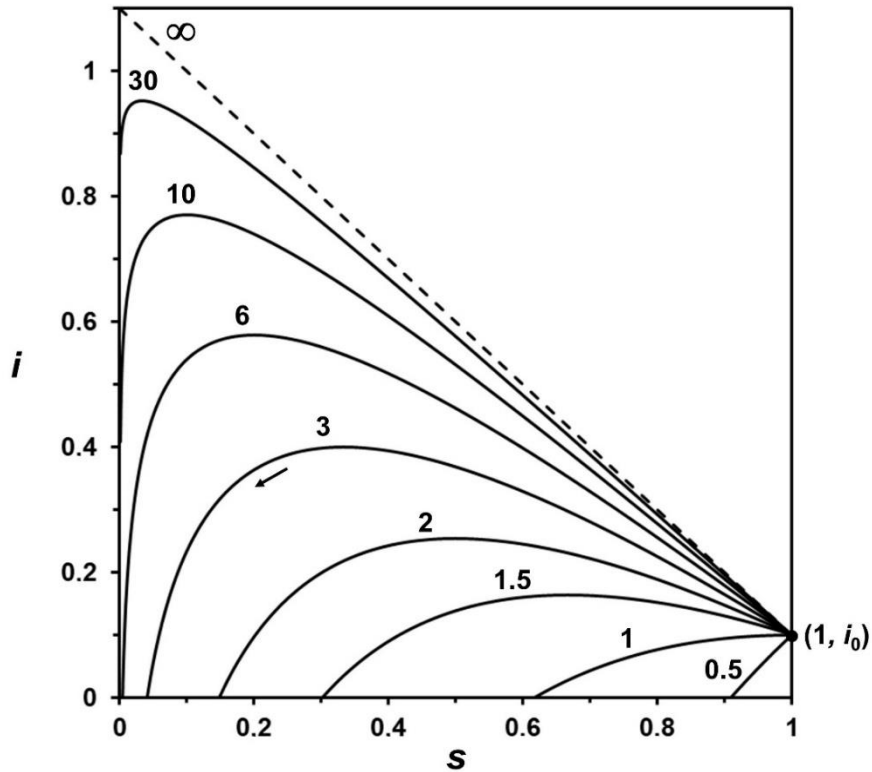

**Fig. 1** The phase space plot: *i* vs. *s* according to equation (20). A value of $i_0 = 0.1$ was used. The number next to each curve is the respective $R_0$. The dashed line ($R_0 = \infty$) corresponds to the logistic curve situation. The arrow of time is also shown.

The maximum value of *i*, $i_c$, occurs for $s = s_c = 1/R_0$, as follows from equation (20), and takes the value (see Figure 1)

$$i_c = 1 + i_0 - \frac{1}{R_0}(1 + \ln R_0). \tag{21}$$

It is useful to solve equation (20) explicitly for *s*. The result is (see Appendix A):

$$s = -\frac{1}{R_0} W_{-1}\left(-R_0 e^{-(1+i_0-i)R_0}\right) \quad \text{if} \quad s \geq s_c, \tag{22}$$

$$s = -\frac{1}{R_0} W_0\left(-R_0 e^{-(1+i_0-i)R_0}\right) \quad \text{if} \quad s \leq s_c, \tag{23}$$

where $W_0(x)$ is the Lambert function computed for the principal branch [11] and $W_{-1}(x)$ is the Lambert function computed for the secondary branch [11]. Using equations (22) and (23), the value of $S$ can be found directly for different points of the epidemic, instead of solving numerically equation (20). In particular, the final value of $s$ (value at the end of epidemic), $s_\infty$, follows from equation (23) by setting $i = 0$,

$$s_\infty = -\frac{1}{R_0} W_0\left(-R_0 e^{-(1+i_0)R_0}\right), \tag{24}$$

as previously obtained from the final size equation [12,13]. Assuming that $i_0 \ll 1$, as is usually the case, equation (24) becomes

$$s_\infty = -\frac{1}{R_0} W_0\left(-R_0 e^{-R_0}\right), \tag{25}$$

For $R_0$ higher but very close to 1, equation (25) reduces to $2/R_0-1$. For large $R_0$, it becomes $\exp(-R_0)$. Instead of equation (24), a simple and accurate formula (average relative error < 1% if $i_0 < 10^{-3}$) and with the correct asymptoptic behaviour can be used,

$$s_\infty = \exp(-R_0) + 2.462\exp(-1.851R_0) + 8.798\exp(-3.580R_0) \quad (R_0 > 1.1). \tag{26}$$

The dependence of $s_\infty$ on $R_0$ is plotted in Figure 2 (see also Figure 1). As discussed above, there is no epidemic for $R_0 < 1$ and the number of susceptibles does not decrease significantly ($i_0 \ll 1$). The fraction of the population that does not catch the disease is about 40% for $R_0 = 1.5$, 20% for $R_0 = 2$, and 6% for $R_0 = 3$. For $R_0 > 5$ virtually the whole (initially susceptible) population is infected.

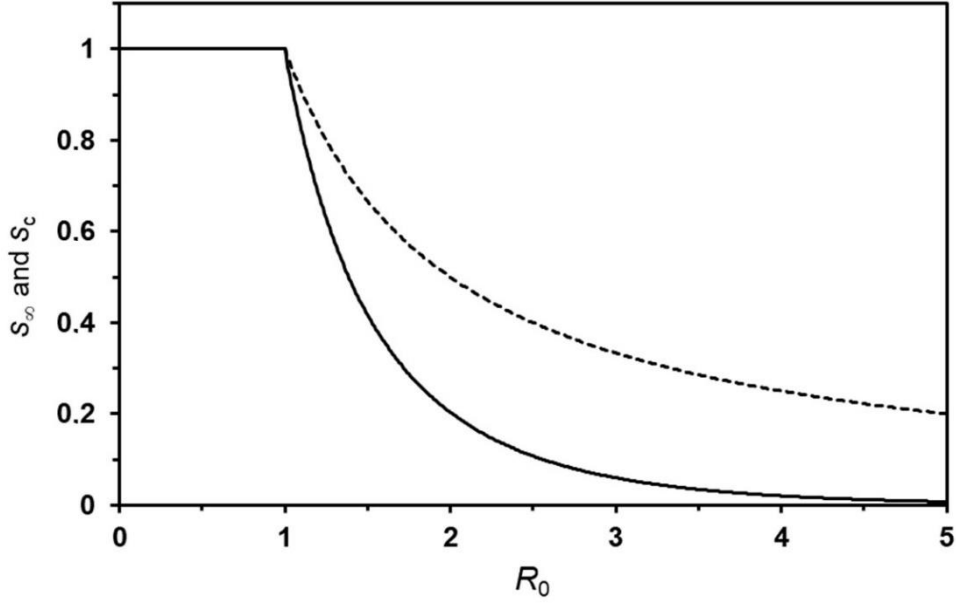

**Fig. 2** The fraction of susceptibles not catching the disease, $s_\infty$ vs. the basic reproduction number $R_0$, as given by equations (25) and (26). It is assumed that $i_0 \ll 1$. Also shown is the critical value $s_c$, equation (18) (dashed line).

Another important parameter that can be obtained is the maximum value of the epidemic curve, $c^*$ (Appendix B):

$$c^* = R_0 s^* \left( s^* - \frac{1}{R_0} \right), \qquad (27)$$

where $s^*$ stands for the number of susceptibles when the production of new infectives is at the maximum, 'epidemic peak' (Appendix B),

$$s^* = -\frac{1}{2R_0} W_{-1}\left(-2R_0 e^{-[1+(1+i_0)R_0]}\right), \qquad (28)$$

where $W_{-1}(x)$ is the Lambert W function computed for the secondary branch. For large $R_0$ equation (27) becomes ($i_0 \ll 1$):

$$c^* = \frac{R_0}{4}. \qquad (29)$$

Finally, the areas under both *i* and *c* curves can be obtained from equations (16) and (14), respectively, and are found to be nearly identical ($i_0 \ll 1$):

$$\int_0^\infty i(\theta)d\theta = 1 + i_0 - s_\infty . \tag{30}$$

$$\int_0^\infty c(\theta)d\theta = 1 - s_\infty . \tag{31}$$

In this way, for large $R_0$ ($R_0 > 5$, say) and very small $i_0$, both *i* and *c* curves are area normalised. The respective peak values, equations (21) and (29), respectively, agree with the fact that for large $R_0$ the curve $i(\theta)$ attains a stable shape (exponential decay function), with a peak value of $1+ i_0$, whereas the curve $c(\theta)$ is increasingly narrow and approaches a delta function.

## 3. Simple functions describing the time-dependence

Obtention of simple closed-form solutions for the time-dependence of the quantities *s*, *i*, *r* and *c* has thus far eluded all efforts, unless $R_0$ is very close to 1 [1-3]. Recently, Barlow and Weinstein [14] presented an accurate closed-form solution using asymptotic approximants. However, the number of terms needed can be quite large and no explicit relations can be obtained for characteristic quantities such as the maximum times.

A different approach is presented here, starting from the limiting results that correspond to $R_0 \to \infty$. In fact, for large $R_0$, one has $R_0 s \gg 1$ until *s* is very close to 0. In this way, virtually all *S* is transformed into *I* before *R* starts to form. Equation (15) can therefore be approximated by

$$\frac{di}{d\theta} = R_0 si , \tag{32}$$

implying that $i + s = i_0 + s_0$ in this time range. The solution of the system of equations (14) and (31) is:

$$s(\theta) = \frac{\dfrac{1}{i_0}+1}{\dfrac{1}{i_0}+\exp\left[(1+i_0)R_0\theta\right]} \quad . \tag{33}$$

The respective $i(\theta)$ is given by the logistic equation [15],

$$i(\theta) = i_0 \frac{1+i_0}{\exp\left[-(1+i_0)R_0\theta\right]+i_0} \quad . \tag{34}$$

Finally, $c(\theta)$ is

$$c(\theta) = R_0\left(1+\frac{1}{i_0}\right)^2 \frac{\exp\left[(1+i_0)R_0\theta\right]}{\left(\dfrac{1}{i_0}+\exp\left[(1+i_0)R_0\theta\right]\right)^2} \quad , \tag{35}$$

and peaks at

$$\theta^* = -\frac{\ln i_0}{(1+i_0)R_0}, \tag{36}$$

with the value

$$c^* = \frac{R_0}{4}(1+i_0)^2, \tag{37}$$

compare equation (29). Equation (36) shows the effect of the initial fraction of infectives on the dynamics. The lower their number, the higher the induction time.

Full time range approximate solutions for the SIR model and for any $R_0$ are now obtained using eq. (33) as the starting point for a trial function representing $s(\theta)$. The trial function is written as

$$s = s_\infty + (1-s_\infty)\frac{a+1}{a+\exp(b\theta)}, \tag{38}$$

where parameters $a$ and $b$ should reduce to $1/i_0$ and to $(1+i_0) R_0$, respectively, in the limit of large $R_0$. The parameter $s_\infty$ is given by equation (24). The only modification in the mathematical form is thus to allow for $s_\infty > 0$. A simple functional form is kept, allowing to analytically compute integrals and characteristic parameters such as those for the maxima of $s$ and $c$. All these quantities can indeed be obtained in closed form from the trial function equation (38). Equation (15) gives

$$\ln\left(\frac{i}{i_0}\right) = \int_0^\theta [R_0 s(u) - 1] du, \qquad (39)$$

and therefore

$$\ln\left(\frac{i}{i_0}\right) = R_0 (1-s_\infty)\left[\frac{\theta}{a} + \frac{1}{b}\left(1+\frac{1}{a}\right)\ln\left(\frac{a+1}{a+e^{b\theta}}\right)\right], \qquad (40)$$

or

$$i = i_0 \left[\frac{(a+1)e^{\gamma\theta}}{a+e^{b\theta}}\right]^\omega e^{-\theta}, \qquad (41)$$

with

$$\gamma = \frac{a+1-s_\infty}{(a+1)(1-s_\infty)} b, \qquad (42)$$

$$\omega = (1-s_\infty)\left(1+\frac{1}{a}\right)\frac{R_0}{b}. \qquad (43)$$

This function has the correct long-time behaviour for large $R_0$, decaying as $\exp(-\theta)$. When $R_0 \to \infty$, parameter $\gamma \to b$, whereas parameter $\omega \to 1$, and the decay of $i$ becomes exponential for all times.

The maximum value of the infectives curve is attained at

$$\theta_c = \frac{1}{b}\ln\left[\frac{R_0-1}{1-R_0 s_\infty}(a+1)\right]. \qquad (44)$$

Note that the alternative computation of $i(\theta)$ from equation (14) is not effective, as $a$ and $b$ are no longer independent and the smoothing effect on deviations by means of integration is nonexistent.

The epidemic curve is obtained from equations (1), (17) and (38),

$$c = -\frac{ds}{dt} = (1-s_\infty)(a+1)b\frac{\exp(b\theta)}{[a+\exp(b\theta)]^2}, \qquad (45)$$

and the time at which the maximum ('epidemic peak') is attained is

$$\theta^* = \frac{\ln a}{b}, \qquad (46)$$

hence the respective peak value is

$$c^* = \frac{(a+1)b(1-s_\infty)}{4a} \simeq \frac{1}{4}b(1-s_\infty) \quad (a \gg 1), \qquad (47)$$

compare equations (29), (36) and (37). In practice, $c(\theta)$ is best computed using the defining equation (17).

Finally, the $r$ curve can in principle be computed either using equation (8) or from

$$r = -\frac{\ln s}{R_0}. \qquad (48)$$

This last relation, obtained by the elimination of time from equations (14) and (16) [3], is less convenient as it amplifies the error of $s$ when this quantity is close to zero.

## 4. Determination of parameters *a* and *b*

Curve fitting for the determination of parameters *a* and *b* (fixed $i_0$ and $R_0$) can be performed in two ways. In the first method, the *s* and *i* curves are obtained by numerical integration, and the fitting is carried out in the time domain in a global way (i.e., minimizing a single sum of squares of deviations and using the same parameters for both curves). In the second method, fitting is carried out in the phase space, no numerical integration being necessary. In this case the exact *i(s)* curve is given by equation (20), and the fitting *i(s)* curve is given by equation (41), the reduced time $\theta$ being computed as a function of *s* with equation (38), rewritten as

$$\theta = \frac{1}{b}\ln\left[1+(a+1)\left(\frac{1-s}{s-s_\infty}\right)\right]. \qquad (49)$$

The dependence of parameters *a* and *b* with $R_0$ for $i_0 = 0.001$ are shown in Figures 3 and 4 (the numerical values are given in the Supplementary Information). Parameter *a* tends to $1/i_0$ when $R_0 \to \infty$, as discussed in connection with equation (38). On the other hand, it is observed that it tends to 1 when $R_0 \to 1$. Parameter *b* follows an almost linear dependence with $R_0$, becoming close to 0 for $R_0 = 1$, as expected. A similar pattern is observed for other values of $i_0$ (up to 0.01; no higher values tested).

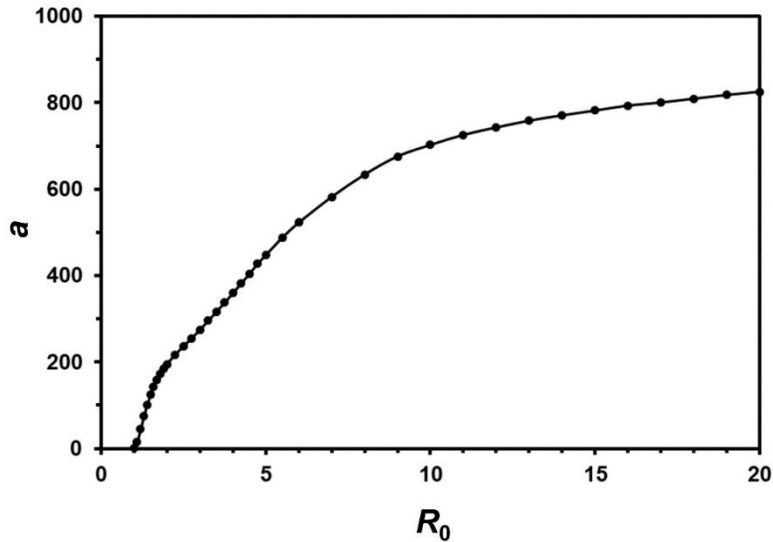

**Fig. 3** Plot of parameter *a* as a function of $R_0$, obtained from the phase space fitting of *i* vs *s* ($i_0 = 0.001$). The initial value is $a(1) = 1$, and the limiting value ($R_0 \to \infty$) is $1/i_0 = 1000$, see equation (38).

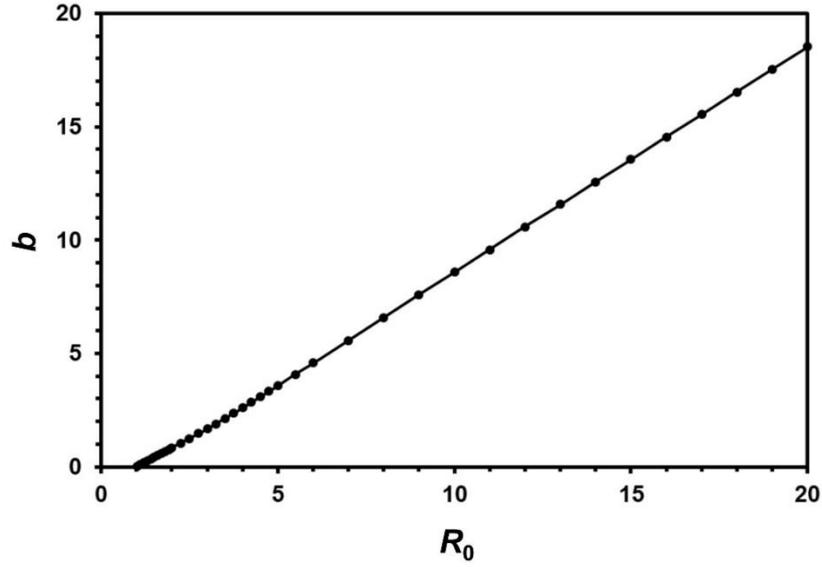

**Fig. 4** Plot of parameter $b$ as a function of $R_0$ obtained from the phase space fitting of $i$ vs $s$ ($i_0 = 0.001$). It is observed that the dependence is almost linear and that $b(1)$ is close to zero.

Fitting for several values of $R_0$ ($1 < R_0 < 1000$) confirms they allow a very good representation of the results. Representative results of phase space fittings are shown in Figures 5-8.

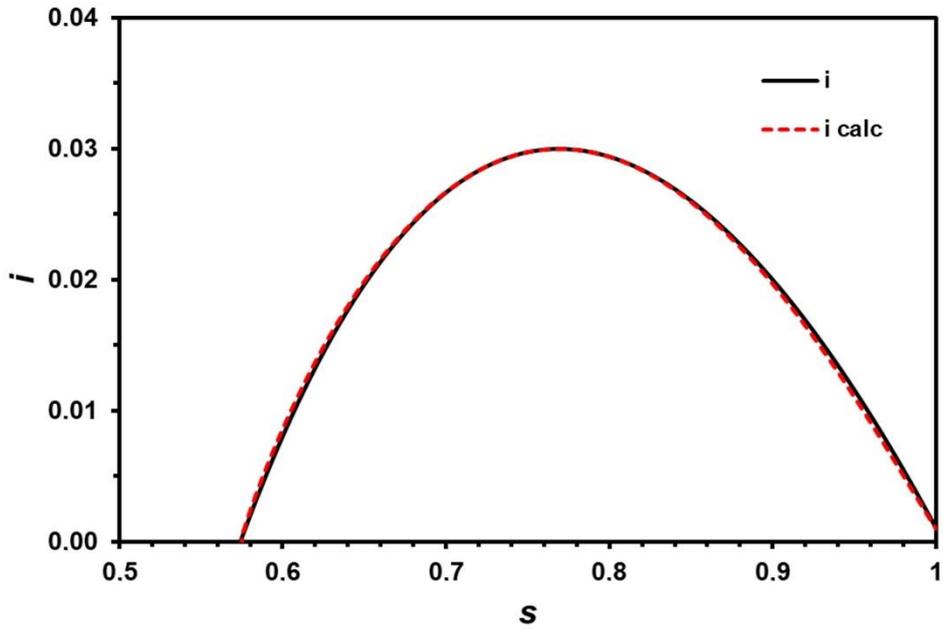

**Fig. 5** Phase space plot for $R_0 = 1.3$ and $i_0 = 0.001$ using equation (20) and the respective fitting using equations (41) and (49). Fitted parameters: $a = 74.40$ and $b = 0.2782$ ($\gamma = 0.6481$ and $\omega = 2.017$).

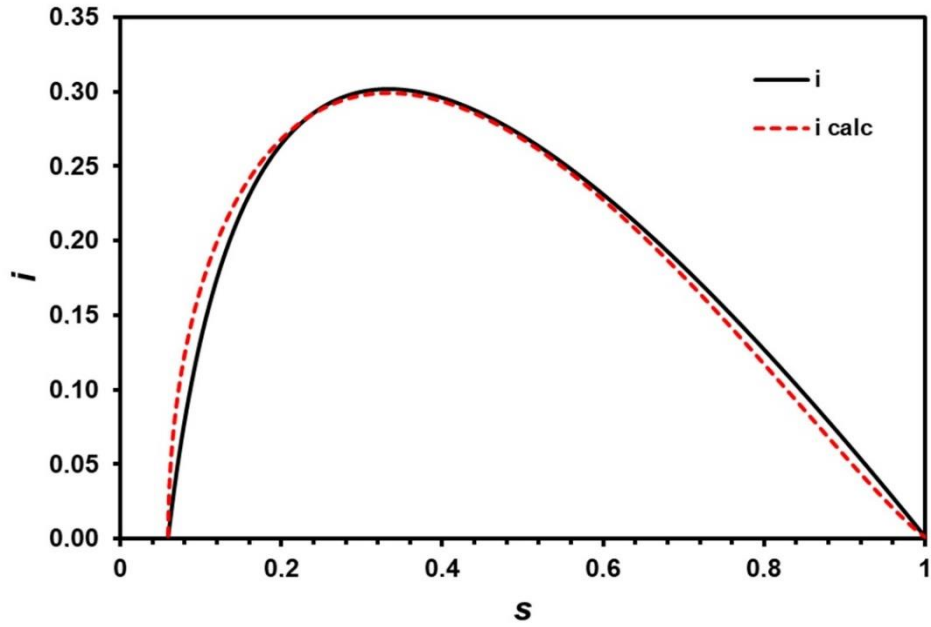

**Fig. 6** Phase space plot for $R_0 = 3.0$ and $i_0 = 0.001$ using equation (20) and the respective fitting using equations (41) and (49). Fitted parameters: $a = 274.5$ and $b = 1.683$ ($\gamma = 1.789$ and $\omega = 1.683$).

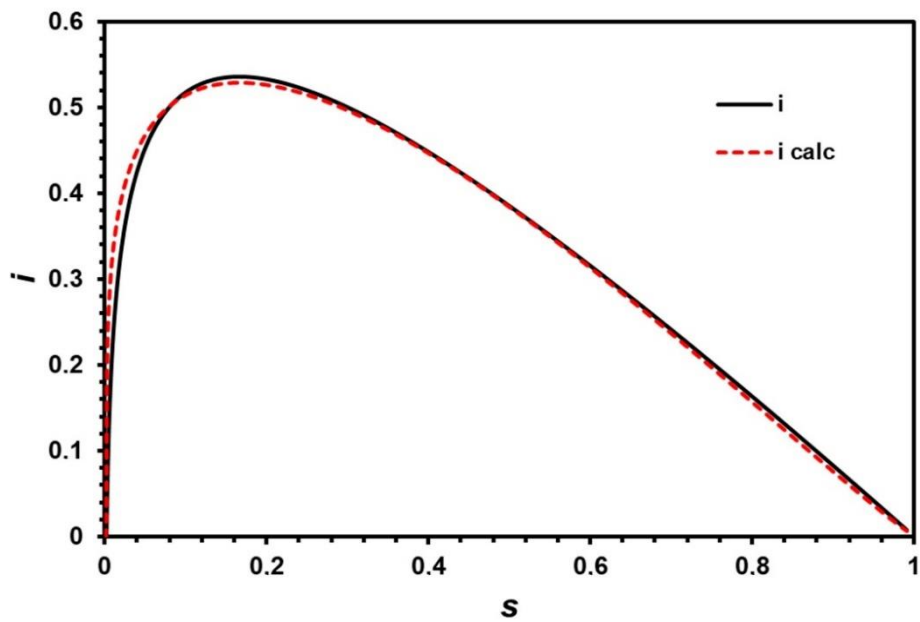

**Fig. 7** Phase space plot for $R_0 = 6.0$ and $i_0 = 0.001$ using equation (20) and the respective fitting using equations (41) and (49). Fitted parameters: $a = 522.8$ and $b = 4.577$ ($\gamma = 4.588$ and $\omega = 1.310$).

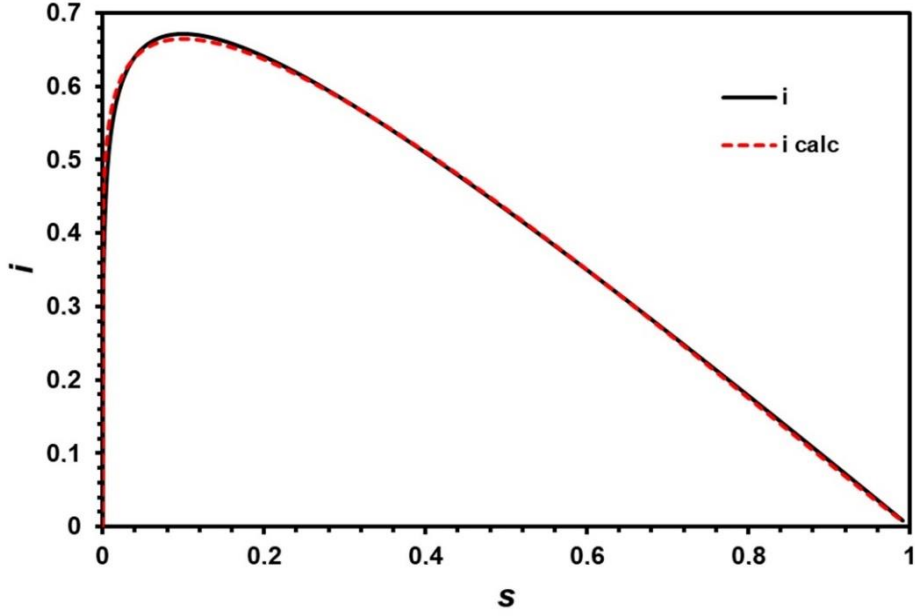

**Fig. 8** Phase space plot for $R_0 = 10.0$ and $i_0 = 0.001$ using equation (20) and the respective fitting using equations (41) and (49). Fitted parameters: $a = 702.7$ and $b = 8.593$ ($\gamma = 8.593$ and $\omega = 1.165$).

Minor deviations for some values of $s$ exist for intermediate values of $R_0$, as seen in Figures 6 and 7. Nevertheless, the overall shapes and the values and position of the maxima are quite well reproduced in all cases. It is possible to use more complicated functions to get better fits for $s$ (and therefore for the remaining quantities), but at the expense of introducing additional parameters and of losing analytical power, for instance

$$s = s_\infty + (1 - s_\infty)\frac{(a+1)\exp(p\theta)}{a + \exp(b\theta)}, \qquad (50)$$

with an additional parameter ($p$), gives very good fits for $s(\theta)$, almost eliminating the somewhat too fast decay observed after the middle point, see e.g. Figures 10 and 11. The chosen set of equations represents a compromise between accuracy and simplicity, allowing to obtain explicit expressions for the peak values and respective times, for instance equations (44), (46) and (47).

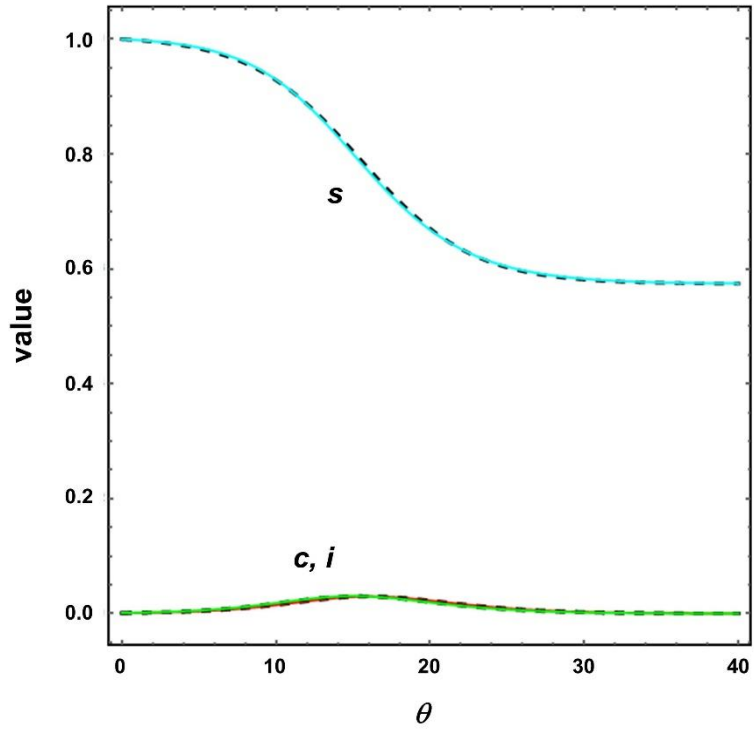

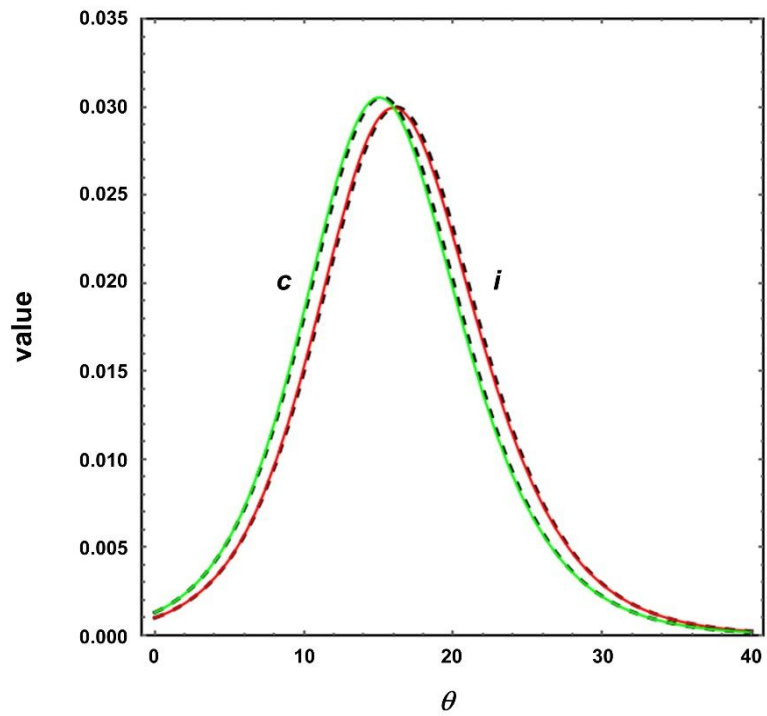

**Fig. 9** Plot of the *s*, *i* and *c* curves as a function of time for $R_0 = 1.3$ and $i_0 = 0.001$, both obtained by numerical integration (color) and by phase space fitting (black, dashed). $\theta^* = 15.4$ and $\theta_c = 16.1$. See also Figure 5.

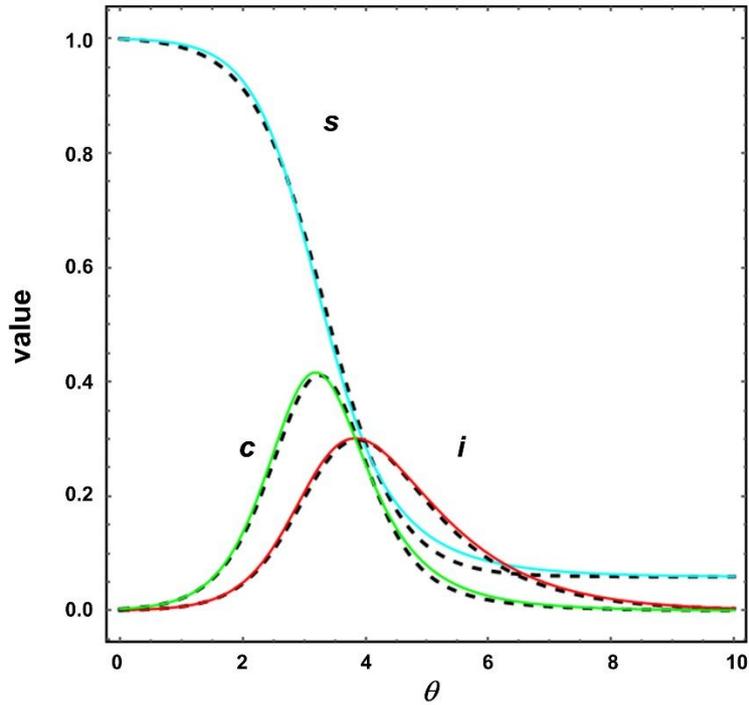

**Fig. 10** Plot of the *s*, *i* and *c* curves as a function of time for $R_0 = 3.0$ and $i_0 = 0.001$, both obtained by numerical integration (color) and by phase space fitting (black, dashed). $\theta^* = 3.34$ and $\theta_c = 3.87$. See also Figure 6.

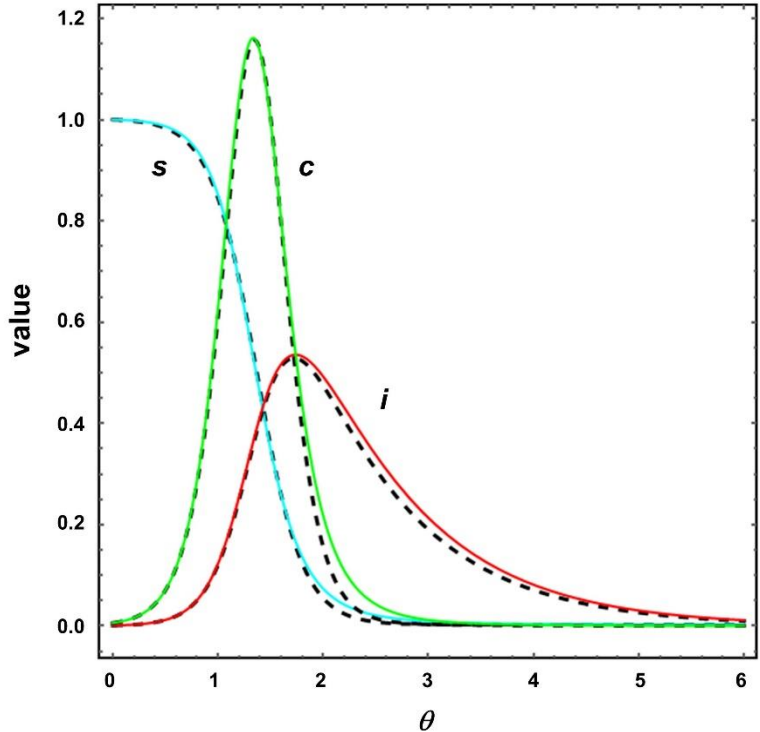

**Fig. 11** Plot of the *s*, *i*, *r* and *c* curves as a function of time for $R_0 = 6.0$ and $i_0 = 0.001$, both obtained by numerical integration (color) and by phase space fitting (black, dashed). $\theta^* = 1.37$ and $\theta_c = 1.72$. See also Figure 7.

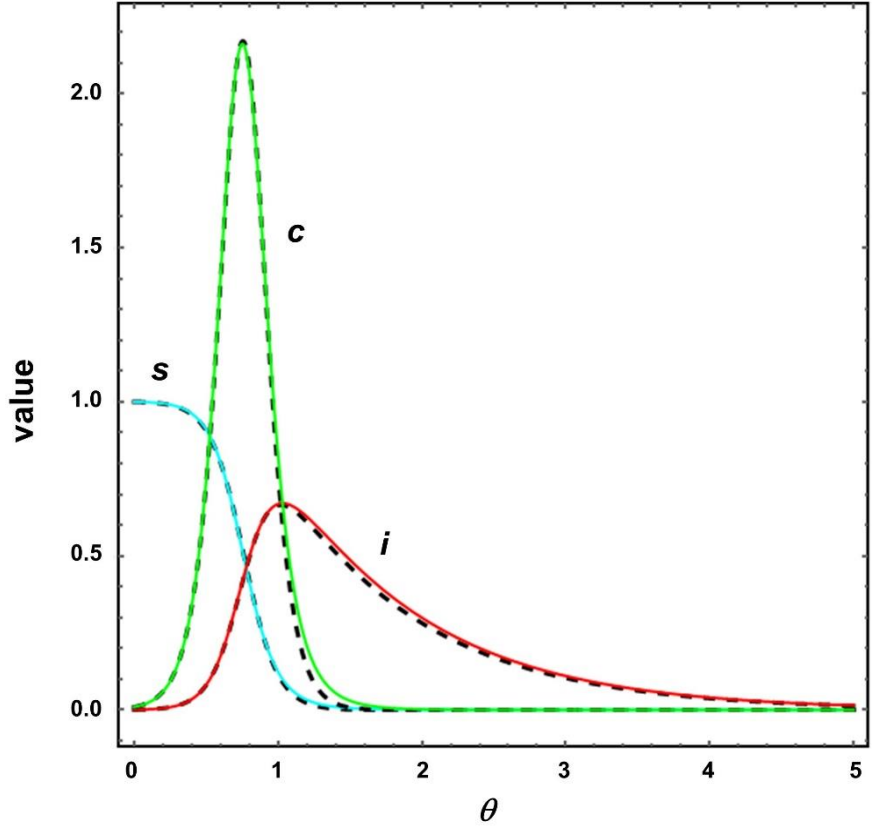

**Fig. 12** Plot of the *s*, *i*, *r* and *c* curves as a function of time for $R_0 = 10.0$ and $i_0 = 0.001$, both obtained by numerical integration (color) and by phase space fitting (black, dashed). $\theta^* = 0.763$ and $\theta_c = 1.02$. See also Figure 8.

Figure 12 clearly demonstrates the impulse-response relationship between $c(\theta)$ and $i(\theta)$, as expressed by equation (7). In particular, after $c(\theta)$ (essentially) ceases, $i(\theta)$ decays exponentially.

## 5. Change in $i_0$: time shift effect

The numerical results presented in the previous section refer to $i_0 = 0.001$. From fittings with other values of $i_0$, it is concluded that parameter *a* has an inverse dependence with $i_0$, for constant $R_0$, whereas parameter *b* practically does not change with $i_0$ ($i_0 < 0.01$), especially for $R_0 > 2$. However, for small $i_0$ ($i_0 < 0.01$), there is no need to evaluate *a* and *b* parameters as a function of $i_0$. The effect of changing this parameter is equivalent to a time shift of the curves, without altering their shape or relative position. Indeed, equation (39) can be rewritten as

$$i(\theta) = i_0 \exp\left\{\int_0^\theta [R_0 s(u) - 1] du\right\}, \tag{51}$$

hence,

$$\begin{aligned} i(\theta) &= i_0 \exp\left\{\int_0^{\Delta\theta} [R_0 s(u) - 1] du\right\} \exp\left\{\int_{\Delta\theta}^\theta [R_0 s(u) - 1] du\right\} = \\ &= i_0 \exp\left\{\int_0^{\Delta\theta} [R_0 s(u) - 1] du\right\} \exp\left\{\int_0^{\theta-\Delta\theta} [R_0 s(u+\Delta\theta) - 1] du\right\}. \end{aligned} \tag{52}$$

Maclaurin series expansion of $s$ gives

$$s(\theta) = 1 - R_0 i_0 \theta + ..., \tag{53}$$

and therefore, for $\Delta\theta \ll 1/(R_0\, i_0)$, one may use $s(\Delta\theta) = s(0) = 1$, and equation (52) becomes

$$\begin{aligned} i(\theta) &= i_0 \exp\left[(R_0 - 1)\Delta\theta\right] \exp\left\{\int_0^{\theta-\Delta\theta} [R_0 s(u+\Delta\theta) - 1] du\right\} = \\ &= i_0' \exp\left\{\int_0^{\theta-\Delta\theta} [R_0 s(u+\Delta\theta) - 1] du\right\} \end{aligned}, \tag{54}$$

where the new initial value, $i_0'$, is

$$i_0' = i_0 \exp\left[(R_0 - 1)\Delta\theta\right]. \tag{55}$$

Comparison of eq. (54) with eq. (51), rewritten as

$$i'(\theta') = i_0' \exp\left\{\int_0^{\theta'} [R_0 s'(u) - 1] du\right\}, \tag{56}$$

gives

$$s'(\theta) = s(\theta + \Delta\theta), \tag{57}$$

and therefore, the time shift relations are obeyed both by $s$ and $i$,

$$s'(\theta') = s(\theta), \tag{58}$$

$$i'(\theta') = i(\theta), \tag{59}$$

where $\theta' = \theta - \Delta\theta$, and the time shift $\Delta\theta$ is

$$\Delta\theta = \frac{1}{R_0 - 1} \ln\left(\frac{i_0'}{i_0}\right). \tag{60}$$

If the $a$ and $b$ parameters for $i_0 = 0.001$ are used, then a time shift is needed for other values of $i_0$. For instance, if $i_0 = 10^{-7}$ and $R_0 = 3$, then the shift (time required to reach $i_0' = 10^{-3}$) is $\Delta\theta = 4.6$, implying an increase of the induction time.

## 6. Conclusions

The SIR epidemic model was formulated in terms of dimensionless variables and parameters, thus reducing the number parameters from four ($S_0$, $I_0$, $\beta$, $\alpha$) to two ($i_0$, $R_0$). The susceptibles population was explicitly related to the infectives population using the Lambert W function (both the principal and the secondary branches), eqs. (22) and (23). A simple and accurate relation for the fraction of the population that does not catch the disease was obtained, eq. (26). The maximum of the epidemic curve (epidemic peak) was also obtained, eq. (27). The explicit time dependences of the susceptibles, infectives (and removed) populations were modelled with good accuracy for any value of $R_0$ using simple functions matching the limiting case $R_0 \to \infty$ (logistic equation), eqs. (38) and (41). From these, the epidemic curve and the peak time are derived, eqs. (45) and (46). It was also shown that for small $i_0$ ($i_0 < 10^{-2}$) the effect of a change in this parameter on the population evolution curves amounts to a time shift, eq. (56), their shape and relative position being unaffected.

**Acknowledgements**

**Competing interests statement**

The author declares that he has no known competing financial interests or personal relationships that could have appeared to influence the work reported in this paper.

**Appendix A: Derivation of equations (22) and (23)**

Equation (20) can be rearranged to give

$$\ln s - R_0 s = -(1+i_0-i)R_0, \qquad (A1)$$

or

$$s e^{-R_0 s} = \exp\left[-(1+i_0-i)R_0\right]. \qquad (A2)$$

Defining $y = -R_0 s$

$$y e^y = -R_0 \exp\left[-(1+i_0-i)R_0\right], \qquad (A3)$$

hence, from the definition of the Lambert W function [11],

$$y = W\left(-R_0 \exp\left[-(1+i_0-i)R_0\right]\right), \qquad (A4)$$

and finally

$$s = -\frac{1}{R_0} W\left(-R_0 \exp\left[-(1+i_0-i)R_0\right]\right). \qquad (A5)$$

Given that the argument of the function is negative, the appropriate branch of W (principal or secondary) must be defined. Let

$$x = -R_0 \exp\left[-(1+i_0-i)R_0\right]. \qquad (A6)$$

Then

$$s = -\frac{1}{R_0} W(x). \qquad (A7)$$

This function is plotted in Figure A1 for $R_0 = 3$ and $i_0 = 0.1$. It is seen that the secondary branch, $W_{-1}(x)$, applies for $t \in [0, t_{max}]$ and the principal branch $W_0(x)$ for $t \in [t_{max}, +\infty[$.

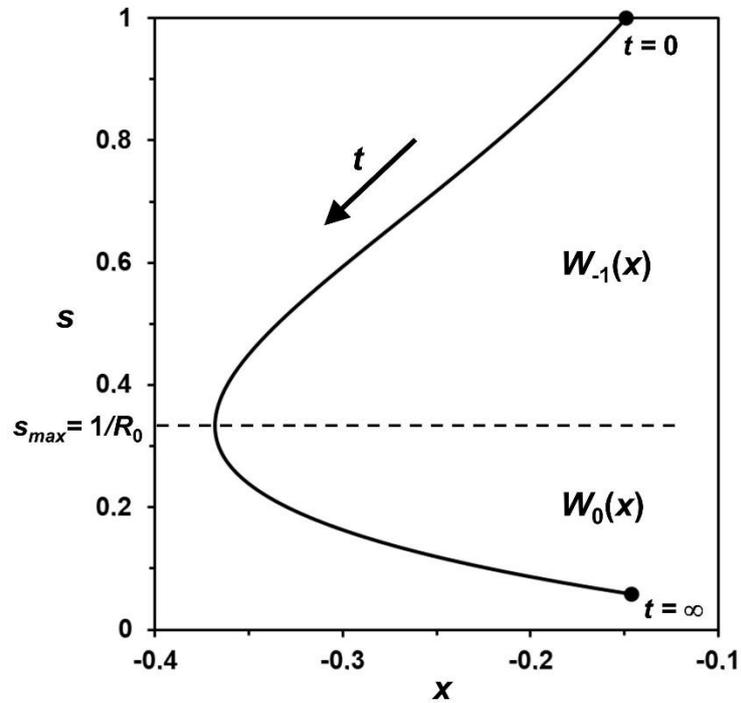

**Fig. A1** Variable $s$ vs $x$, for $R_0 = 3$ and $i_0 = 0.1$, showing the two branches of $W(x)$. $s_{max}$ represents the value of $s$ when $i$ attains its maximum value, $i_{max}$.

**Appendix B: Derivation of equations (27) and (28)**

The rate of production of new infectives (*epidemic curve*) is $-ds/d\theta$, and the maximum rate occurs when

$$\frac{d^2 s}{d\theta^2} = 0. \tag{B1}$$

It follows from equations (B1), (14) and (15) that

$$i^* = s^* - 1/R_0 = s^* - s_c, \tag{B2}$$

where $s^*$ and $i^*$ stand for the number of susceptibles and infectives when the production of new infectives is at the maximum ($c^*$).

The maximum rate is thus

$$c^* = R_0 s^* \left( s^* - \frac{1}{R_0} \right). \tag{B3}$$

Using equations (B2) and (20) it is obtained that

$$2s^* - \frac{1}{R_0} \ln s^* - \left( 1 + i_0 + \frac{1}{R_0} \right) = 0. \tag{B4}$$

It follows from equation (B2) that $s^* > s_c$. The solution of equation (B4) is then

$$s^* = -\frac{1}{2R_0} W_{-1}\left( -2R_0 e^{-[1+(1+i_0)R_0]} \right), \tag{B5}$$

where $W_{-1}(x)$ is the Lambert W function computed for the secondary branch (see Appendix A).

The dependence of $s^*$ with $R_0$ ($i_0 \ll 1$) is plotted in Figure B1. The reduced parameter starts from $s^* = 1$ for $R_0 = 1$, but quickly approaches the asymptotic value $1/2$ as $R_0$ increases and is reasonably constant for $R_0 > 5$.

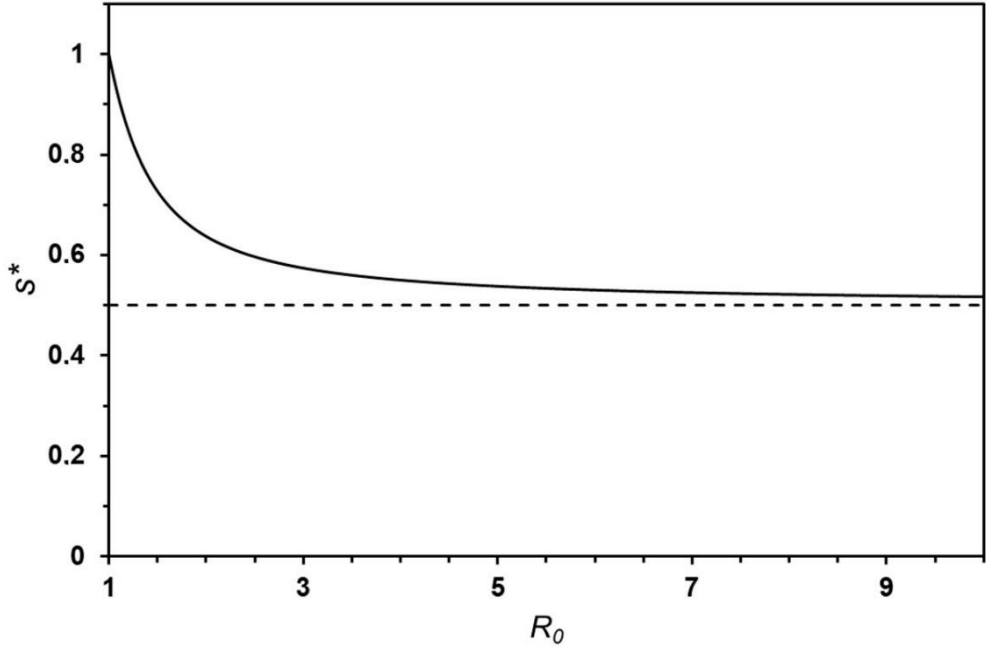

**Fig. B1** The dependence of $s^*$ with $R_0$ ($i_0 \ll 1$).

A simple formula, with an accuracy better than 0.1%, and with the correct asymptoptic behaviour is ($i_0 \ll 1$):

$$s^* = 3.40\exp(-2.17R_0) + 0.143\exp(-0.247R_0) + \frac{1}{2}. \qquad (B6)$$

Given that $s^*$ approaches 1/2 for large $R_0$, it follows from equation (B3) that $c^*$ also tends to a constant value for high $R_0$ ($i_0 \ll 1$),

$$c^* = \frac{R_0}{4}. \qquad (B7)$$

Indeed, for large $R_0$ the logistic curve applies in this time range and $s$ and $i$ cross at $s^* = i^* = 1/2$.

# Supplementary information

**Recovered *a* and *b* parameters and related quantities as a function of $R_0$ ($i_0 = 0.001$)**

| $R_0$ | *a* | *b* | $s_\infty$ | $\gamma$ | $\omega$ | $\theta_c$ | $\theta*$ |
|---|---|---|---|---|---|---|---|
| 1.01 | 1.477324 | 0.045688 | 0.945215 | 0.515759 | 2.030899 | - | 8.541242 |
| 1.1 | 16.07543 | 0.105981 | 0.814742 | 0.544778 | 2.042442 | 26.42449 | 26.2055 |
| 1.2 | 44.47559 | 0.192416 | 0.681721 | 0.595489 | 2.029573 | 20.33014 | 19.72258 |
| 1.3 | 74.39597 | 0.278187 | 0.574059 | 0.648139 | 2.017226 | 16.14126 | 15.49102 |
| 1.4 | 101.339 | 0.362169 | 0.486834 | 0.702397 | 2.003268 | 13.40905 | 12.75225 |
| 1.5 | 124.0802 | 0.444453 | 0.415525 | 0.757905 | 1.988462 | 11.50196 | 10.84688 |
| 1.6 | 143.0793 | 0.525651 | 0.356684 | 0.815073 | 1.971841 | 10.09249 | 9.442391 |
| 1.7 | 158.8413 | 0.606022 | 0.307713 | 0.873706 | 1.954215 | 9.006243 | 8.362579 |
| 1.8 | 172.0259 | 0.68587 | 0.266644 | 0.933807 | 1.935811 | 8.141698 | 7.505279 |
| 1.9 | 185.5023 | 0.767833 | 0.232448 | 0.99912 | 1.909543 | 7.431112 | 6.802346 |
| 2 | 195.3889 | 0.847624 | 0.20292 | 1.062313 | 1.890365 | 6.843481 | 6.223265 |
| 2.25 | 217.172 | 1.049542 | 0.146436 | 1.228775 | 1.838289 | 5.724527 | 5.126702 |
| 2.5 | 236.5981 | 1.255507 | 0.107303 | 1.405784 | 1.785077 | 4.928983 | 4.35391 |
| 2.75 | 255.1503 | 1.466131 | 0.079552 | 1.592349 | 1.733238 | 4.332671 | 3.779917 |
| 3 | 274.4621 | 1.682726 | 0.05952 | 1.788834 | 1.682817 | 3.867708 | 3.336736 |
| 3.25 | 295.5264 | 1.906333 | 0.044859 | 1.995565 | 1.633875 | 3.493958 | 2.984136 |
| 3.5 | 316.2948 | 2.133682 | 0.034011 | 2.208569 | 1.589576 | 3.188321 | 2.698 |
| 3.75 | 337.4099 | 2.365566 | 0.025912 | 2.428306 | 1.548745 | 2.932947 | 2.460848 |
| 4 | 359.3582 | 2.60214 | 0.01982 | 2.654611 | 1.510922 | 2.716345 | 2.261339 |
| 4.25 | 382.0144 | 2.843031 | 0.01521 | 2.886827 | 1.476 | 2.530242 | 2.09124 |
| 4.5 | 403.9778 | 3.085961 | 0.011704 | 3.122416 | 1.444717 | 2.369019 | 1.94473 |
| 4.75 | 427.2675 | 3.333652 | 0.009026 | 3.363945 | 1.415308 | 2.227384 | 1.817049 |
| 5 | 447.6966 | 3.579785 | 0.006974 | 3.604868 | 1.39009 | 2.102956 | 1.705163 |
| 5.5 | 488.097 | 4.078617 | 0.00418 | 4.095702 | 1.345611 | 1.892773 | 1.517798 |
| 6 | 522.7816 | 4.576859 | 0.002516 | 4.58838 | 1.310146 | 1.722955 | 1.367568 |
| 7 | 582.4462 | 5.57831 | 0.000918 | 5.583425 | 1.255861 | 1.464092 | 1.141428 |
| 8 | 634.2255 | 6.588663 | 0.000336 | 6.590877 | 1.215712 | 1.27531 | 0.979319 |
| 9 | 676.1828 | 7.598492 | 0.000124 | 7.599429 | 1.186051 | 1.131606 | 0.8576 |
| 10 | 702.7105 | 8.592561 | $4.54\times10^{-5}$ | 8.592951 | 1.165401 | 1.018794 | 0.762863 |
| 11 | 725.9055 | 9.589276 | $1.67\times10^{-5}$ | 9.589436 | 1.148676 | 0.927241 | 0.686957 |
| 12 | 743.1268 | 10.5807 | $6.14\times10^{-6}$ | 10.58076 | 1.13566 | 0.851568 | 0.624804 |
| 13 | 758.6851 | 11.57449 | $2.26\times10^{-6}$ | 11.57452 | 1.124637 | 0.787753 | 0.572948 |
| 14 | 771.1433 | 12.56552 | $8.32\times10^{-7}$ | 12.56553 | 1.115604 | 0.733287 | 0.529057 |
| 15 | 782.1352 | 13.5574 | $3.06\times10^{-7}$ | 13.55741 | 1.107821 | 0.686147 | 0.491394 |
| 16 | 793.7616 | 14.55469 | $1.13\times10^{-7}$ | 14.55469 | 1.100687 | 0.644885 | 0.458738 |
| 17 | 800.7273 | 15.54165 | $4.14\times10^{-8}$ | 15.54165 | 1.095201 | 0.608646 | 0.430168 |
| 18 | 809.7036 | 16.53771 | $1.52\times10^{-8}$ | 16.53771 | 1.089766 | 0.576326 | 0.404933 |
| 19 | 818.2248 | 17.53533 | $5.60\times10^{-9}$ | 17.53533 | 1.084851 | 0.547394 | 0.382493 |
| 20 | 825.7082 | 18.53218 | $2.06\times10^{-9}$ | 18.53218 | 1.080511 | 0.521357 | 0.36241 |

| 50 | 918.0009 | 48.47956 | $1.93 \times 10^{-22}$ | 48.47956 | 1.032486 | 0.221023 | 0.140723 |
| 100 | 958.1853 | 98.52386 | $3.72 \times 10^{-44}$ | 98.52386 | 1.016042 | 0.116329 | 0.069679 |
| 150 | 970.6981 | 148.5439 | $7.18 \times 10^{-66}$ | 148.5439 | 1.010843 | 0.079997 | 0.046303 |
| 200 | 976.3637 | 198.5443 | $1.38 \times 10^{-87}$ | 198.5443 | 1.008363 | 0.061337 | 0.034672 |
| 500 | 989.2354 | 498.748 | $7.10 \times 10^{-218}$ | 498.748 | 1.003524 | 0.026287 | 0.013829 |
| 1000 | 994.139 | 999.177 | 0 | 999.177 | 1.00183 | 0.013821 | 0.006908 |